\documentclass[12pt]{article}
\usepackage{amsmath,amssymb,amsfonts,amsthm,amscd}
\usepackage{graphicx,psfrag,epsfig}
\usepackage[T1]{fontenc}
\usepackage[utf8]{inputenc}

\usepackage[font=small,labelfont=bf,width=\textwidth]{caption}

\newtheorem{theorem-definition}[theorem]{Theorem-Definition}
\newtheorem{theorem-construction}[theorem]{Theorem-Construction}
\newtheorem{lemma-definition}[theorem]{Lemma--Definition}
\newtheorem{lemma-construction}[theorem]{Lemma--Construction}

\let\oldbfseries=\bfseries
\let\oldmdseries=\mdseries
\let\oldnormalfont=\normalfont
\renewcommand{\bfseries}{\oldbfseries\boldmath}
\renewcommand{\mdseries}{\oldmdseries\unboldmath}
\renewcommand{\normalfont}{\oldnormalfont\unboldmath}

\allowdisplaybreaks[3]

\numberwithin{equation}{section}

\makeatletter
\def\blfootnote{\xdef\@thefnmark{}\@footnotetext}
\makeatother

\def\nn{\nonumber}
\def\nnn{\nonumber\\}
\def\pa{\partial}

\newcommand{\be}{\begin{equation}}
\newcommand{\ee}{\end{equation}}

\def\({\left(}
\def\){\right)}

\renewcommand{\[}{\left[}
\renewcommand{\]}{\right]}
\def\blb{\bigl[}
\def\brb{\bigr]}
\def\lcurl{\left\{}
\def\rcurl{\right\}}
\def\blcurl{\big\{}
\def\brcurl{\big\}}

\def\a{\alpha}
\def\b{\beta}
\def\g{\gamma}
\def\d{\delta}
\def\D{\Delta}

\def\k{\kappa}
\def\l{\lambda}
\def\m{\mu}

\def\tPhi{\widetilde{\Phi}}
\def\tF{\widetilde{F}}
\def\tG{\widetilde{G}}

\def\bpsi{\bar{\psi}}
\def\bchi{\bar{\chi}}

\def\Thp{\Theta_{+}}
\def\Thm{\Theta_{-}}
\def\Thpm{\Theta_{\pm}}

\def\lie{\mathcal{G}}
\def\alie{{\widehat\lie}}

\def\cm{{\cal M}}
\def\ck{{\cal K}}
\def\U{U}
\def\V{V}

\def\H#1{H^{#1}}
\def\Ea#1{E_{+}^{#1}}
\def\Eb#1{E_{-}^{#1}}
\def\Eab#1{E_{\pm}^{#1}}
\def\D#1{D^{(#1)}}
\def\hc{\hat{c}}
\def\hd{\hat{d}}

\def\integer{\mathbb{Z}}

\def\bra#1{\langle#1|}
\def\ket#1{|#1\rangle}

\begin{document}

\title{The higher grading structure of the WKI
hierarchy \\ and the two-component short pulse equation}

\author{G. S. Fran\c ca${}^*$, J. F. Gomes${}^\dagger$, 
A. H. Zimerman${}^\ddagger$}%
\blfootnote{${}^*$guisf@ift.unesp.br, ${}^\dagger$jfg@ift.unesp.br, 
${}^\ddagger$zimerman@ift.unesp.br}

\date{May, 2012}

\maketitle

\vspace{-1.2cm}

\begin{center}
{\it\small Instituto de F\' isica Te\' orica - IFT/UNESP\\
Rua Dr. Bento Teobaldo Ferraz, 271, Bloco II\\
01140-070, S\~ ao Paulo - SP, Brazil}
\end{center}

\begin{abstract}
A higher grading affine algebraic construction of integrable hierarchies,
containing the Wadati-Konno-Ichikawa (WKI) hierarchy as a particular 
case, is proposed. We show that a two-component generalization of the 
Sch\" afer-Wayne short pulse equation arises quite naturally from the 
first negative flow of the WKI hierarchy. Some novel integrable nonautonomous 
models are also proposed.
The conserved charges, both local and nonlocal, are obtained from
the Riccati form of the spectral problem.
The loop-soliton solutions of the WKI hierarchy are systematically constructed
through gauge followed by reciprocal B\" acklund transformation, establishing 
the precise connection between the whole WKI and AKNS hierarchies. 
The connection between the short pulse equation with the sine-Gordon
model is extended to a correspondence between the two-component short pulse 
equation and the Lund-Regge model.
\end{abstract}

\section{Introduction}
\label{sec:intro}
The classification of integrable models is of fundamental importance.
A systematization in this vein is the graded affine algebraic construction 
of integrable hierarchies \cite{Aratyn_construction,Aratyn_symmetry_flows},
unifying in a single algebraic 
structure, different models that at first sight appear unrelated.
For instance, the modified Korteweg-de Vries (mKdV) 
and sine/sinh-Gordon models are members of the same hierarchy. Another 
example is given by the cubic nonlinear Schr\" odinger (NLS) and 
the Lund-Regge models, which are members of the 
Ablowitz-Kaup-Newell-Segur (AKNS) hierarchy. Due to the fact that
all the models within a hierarchy share 
the same algebraic structure, solitonic solutions for all of them
can be constructed in a universal manner, through the dressing 
method \cite{Babelon_dressing}.

The short pulse equation (SPE)
\begin{equation}\label{short_pulse}
\tfrac{1}{4}u_{xt} = u + \tfrac{1}{6}\(u^3\)_{xx}
\end{equation}
was proposed \cite{Schafer_spe} as a model to describe ultra-short 
optical pulses traversing within a nonlinear media. This equation replaces 
the standard NLS equation in the ultra-short pulse regime, being a very good 
approximation of Maxwell's equations.
Since then, \eqref{short_pulse}  was proved to be integrable 
\cite{Sakovich_spe_integrable} having a Lax pair 
of the Wadati-Konno-Ichikawa (WKI) type \cite{Wadati_wki}. Moreover,
it was shown \cite{Sakovich_spe_integrable} that \eqref{short_pulse} is 
related, through a hodograph transformation, to the sine-Gordon equation
\begin{equation}\label{sine_gordon}
\theta_{ys} = 4\sin\theta
\end{equation}
from whence its solitary solutions were first obtained 
\cite{Sakovich_solitary_spe}.
The hodograph transformation was further explored 
\cite{Matsuno_spe_periodic} to 
construct solitonic and quasi-periodic solutions. The transformation
connecting \eqref{short_pulse} and \eqref{sine_gordon} reads 
\begin{equation}\label{hodograph}
\begin{split}
dx &= \a dy - 2u^2 dt, \\
dt &= ds, \\
\a^2 &= \(1+u_x^2\)^{-1} = \cos\theta, \\
u &= \tfrac{1}{4}\theta_s.
\end{split}
\end{equation}
Through a recursion operator approach, it was proposed in
\cite{Brunelli_spe_hierarchy} a hierarchy containing \eqref{short_pulse} and 
also the 
elastic beam equation (EBE) \cite{Ichikawa_ebe}
\begin{equation}\label{elastic_beam}
u_{xt} = \dfrac{1}{4}\(\dfrac{u_{xx}}{\(1+u_x^2\)^{3/2}}\)_{xx}.
\end{equation}
The bi-Hamiltonian structure and nonlocal conservation charges for this 
single component field hierarchy 
were also proposed \cite{Brunelli_spe_bihamiltonian}.
Recently, a multi-component generalization of the short pulse
equation was proposed \cite{Matsuno_nspe} together with its soliton solutions.

On the other hand, the algebraic construction 
\cite{Aratyn_construction, Aratyn_symmetry_flows} of integrable 
hierarchies is based on Toda field theories and the fields appear 
as coefficients of a linear combination of \emph{zero grade} operators.
This construction was further generalized 
\cite{Gervais_higher_grading,Ferreira_affine_toda_matter,Assis_bullough_dodd} 
by the addition of fields associated to \emph{higher grading} operators,
yielding the generalized affine Toda models. The higher grade fields are
physically interpreted as matter fields with the usual Toda fields
coupled to them.

In this paper, we propose a general \emph{higher grading} construction for 
the zero curvature equation, containing the WKI hierarchy as a particular case.
In our construction, the zero grade Toda fields are completely removed, 
remaining the higher grade fields only. 
Considering the WKI hierarchy, its previous known models appear as 
members of the positive (time) flows. We then extend the WKI hierarchy to 
incorporate negative flows, which were not been previously considered. 
Deriving the first negative flow, a model which is a two-component 
generalization of the SPE \eqref{short_pulse} emerges quite 
naturally, showing that the SPE is in fact a natural member 
of the WKI hierarchy and underlies the same algebraic structure as the 
previous known models within the positive flows.
This two-component short pulse equation (2-SPE) was also introduced 
recently in \cite{Matsuno_nspe} under a different context. 
We also introduce some 
novel nonautonomous models, mixing a positive flow with a negative one
\cite{Gomes_mixed}. 
The conserved charges, both local and nonlocal, are constructed from 
the Riccati form of the spectral problem \cite{Wadati_conservation}. 

The WKI and AKNS hierarchies are known to be related 
\cite{Ishimori_akns_wki,Rogers_reciprocal}, but this connection was
not sufficiently explored. Driven by this results we establish, 
from an algebraic perspective, the relation between both hierarchies, thus 
systematizing and extending the hodograph transformation \eqref{hodograph} 
for the whole WKI hierarchy. 
As an example, we explicitly derive a correspondence between 
the 2-SPE and the Lund-Regge model, where \eqref{hodograph} appear as a 
particular case.
Our relations also provide explicit and
systematic solutions for all the models within the WKI hierarchy in terms
of the known AKNS tau functions.

We organize our work as follows. In section \ref{sec:usual} we 
introduce the main algebraic concepts and the usual construction. 
In section \ref{sec:higher}, a higher 
grading integrable hierarchy with an arbitrary affine Lie algebra $\alie$
is proposed, then, in section \ref{sec:wki}, 
this construction yields the WKI hierarchy by choosing $\alie=\hat{A}_1$ with 
homogeneous gradation. The known models of the WKI hierarchy are derived
by considering the positive flows and the 2-SPE emerges from the first 
negative flow. We also propose some novel nonautonomous integrable 
models. 
In section \ref{sec:conservation}, we derive the local and nonlocal 
conserved charges for the WKI hierarchy. In section \ref{sec:dressing}, we 
show that the dressing method is unable to solve the WKI hierarchy, then
we gauge the AKNS hierarchy into the WKI hierarchy with the aid of a reciprocal 
B\" acklund transformation, mapping the corresponding flows of both 
hierarchies. 
We consider explicitly this mapping between the 
first negative flows, obtaining a correspondence between the 2-SPE 
and the Lund-Regge model. Our final remarks are presented in
section \ref{sec:conclusions}.

\section{The usual algebraic construction}
\label{sec:usual}
Let $\alie$ be an affine Kac-Moody algebra and $Q$ an operator 
decomposing the algebra into the graded subspaces
\begin{equation}\label{zgrading}
\alie = \bigoplus_{j\in\integer} \alie^{(j)}, \qquad 
\blb Q, \alie^{(j)} \brb = j \alie^{(j)}.
\end{equation}
As a consequence of the Jacobi identity, 
$\blb \alie^{(i)}, \alie^{(j)} \brb \subset \alie^{(i+j)}$.
Let $E$ be a semi-simple element, with a definite grade, defining the kernel 
subspace
\begin{equation}
\ck \equiv \blcurl T \in \alie \ \vert \ \blb E, T \brb = 0 \brcurl.
\end{equation}
The image subspace, $\cm$, is its complement and $\alie=\ck\oplus\cm$.
Then we have the relations $\blb \ck, \ck \brb \subset \ck$, 
$\blb \ck,\cm \brb \subset \cm$ and we assume the symmetric space structure 
$ \blb \cm, \cm \brb \subset \ck$.

Integrable hierarchies are constructed from the zero curvature
equation
\begin{equation}\label{zero_curvature}
\blb \pa_x + \U, \pa_t + \V \brb = 0,
\end{equation}
where the Lax pair $\U$ and $\V$ lie on $\alie$ and have the following form,
\begin{equation}\label{usual_const}
\U \equiv E^{(1)} + A^{(0)}[\phi]
\end{equation}
where $E=E^{(1)}\in\alie^{(1)}$ is a constant semi-simple element, 
defining $\ck$ and $\cm$, and 
$A^{(0)}[\phi] \in \cm^{(0)}$, where $\cm^{(0)} = \cm \cap \alie^{(0)}$,
is the operator containing the fields of the models. For instance,
if $\cm^{(0)}$ is spanned by $\blcurl T^{(0)}_1,\dotsc,T^{(0)}_h \brcurl$ 
then $A^{(0)}[\phi]=\phi_1 T^{(0)}_1+\dotsb+\phi_h T^{(0)}_h$, 
$\phi_i= \phi_i\(x,t\)$ and $\phi = \(\phi_1,\dotsc,\phi_h\)^T$.
The other Lax operator has the form
\begin{align}
\V &\equiv \sum_{i=0}^{n}\D{i} \qquad \textnormal{(for positive flows)}, \\ 
\V &\equiv \sum_{i=-n}^{-1}\D{i} \qquad \textnormal{(for negative flows)}.
\end{align}
The positive integer $n$ labels the (time) flow under quest.
Formally, we should have written $\V=\V\(n\)$ and $t=t_{n}$, showing 
explicitly the flow we are considering, i.e. the particular model within the
hierarchy, but we usually omit the index $n$ 
that will be clearly indicated from the context.
It is possible to show that the operators $\D{i}\in\alie^{(i)}$ are 
determined in terms of the fields contained in $A^{(0)}$ by projecting 
the zero curvature equation \eqref{zero_curvature} into each graded 
subspace according to \eqref{zgrading}. This procedure enables us to
construct the operator $\V$, one for each flow $n$, in terms of the
algebraic structure of the operator $\U$, thus yielding a nonlinear 
model together with its Lax pair.
Therefore, by construction, all the models are immediately classified 
through $\langle \alie, Q, E^{(1)} \rangle$, defining 
an integrable hierarchy of differential equations.

\section{A higher grading construction}
\label{sec:higher}
We now propose a higher grading construction without the usual Toda fields,
keeping the higher grade fields only.  
Let us define the following zero curvature Lax pair
\begin{subequations}\label{operators}
\begin{align}
U &\equiv E^{(1)}+A^{(1)}[\phi], \label{operators_u}\\
V &\equiv \sum_{i=-m}^{n}\D{i}[\phi], \label{operators_v}
\end{align}
\end{subequations}
where $E=E^{(1)}\in\alie^{(1)}$ is a constant semi-simple element 
and $A^{(1)}[\phi]\in\cm^{(1)}$, where $\cm^{(1)}=\cm\cap\alie^{(1)}$, is 
the operator containing the fields now having \emph{grade one}. 
The positive integers $n$ and $m$ 
label each model, mixing a positive flow $n$ with a negative 
flow $m$ \cite{Gomes_mixed}.
To study positive and negative flows separately one should consider
\begin{align}
V &\equiv
\sum_{i=1}^{n}\D{i}\qquad\text{(for positive flows)},\label{pos_flow} \\
V &\equiv
\sum_{i=-n}^{1}\D{i}\qquad \text{(for negative flows)}. \label{neg_flow}
\end{align}
With this algebraic structure, the zero curvature equation can be solved
non trivially in the following way.
Taking \eqref{zero_curvature} with \eqref{operators}, the
projection into each graded subspace yields the following set of equations
\begin{equation}\label{set_equations}
\begin{split}
\blb E^{(1)}+A^{(1)},\D{n}\brb &= 0 \qquad \text{grade $n+1$}, \\
\pa_x\D{n}+\blb E^{(1)}+A^{(1)},\D{n-1}\brb &= 0 \qquad \text{grade $n$},\\
& \ \, \vdots \\
\pa_x\D{1} - \pa_t A^{(1)}+\blb E^{(1)}+A^{(1)},\D{0}\brb &= 0 
\qquad \text{grade $1$},  \\
& \ \, \vdots  \\
\pa_x\D{-m+1}+\blb E^{(1)}+A^{(1)},\D{-m}\brb&=0 \qquad \text{grade $-m+1$},\\
\pa_x\D{-m} &= 0 \qquad \text{grade $-m$}.
\end{split}
\end{equation}
Note that the equations of motion comes from the projection into 
$\cm\cap\alie^{(1)}$.
Each equation is further decomposed into $\ck$ and $\cm$ components.
For mixed or negative flows,
we have the equations of motion in the following form
\begin{equation}\label{field_eq_neg}
\pa_t A^{(1)} = \pa_x \D{1}_{\cm} + \blb E^{(1)}, \D{0}_{\cm} \brb +
\blb A^{(1)}, \D{0}_{\ck} \brb,
\end{equation}
while for positive flows the equations of motion are given by
\begin{equation}\label{field_eq_pos}
\pa_t A^{(1)} = \pa_x \D{1}_{\cm},
\end{equation}
where $\D{j} = \D{j}_{\ck}+\D{j}_{\cm}$.  
We remark that this construction is valid for an arbitrary 
affine Lie algebra $\alie$ with a grading operator $Q$ and arbitrary grade 
one semi-simple element $E^{(1)}$.

\section{The Wadati-Konno-Ichikawa hierarchy}
\label{sec:wki}
Let us now consider the Kac-Moody algebra $\hat{A}_1=\blcurl \H{j}, \Ea{j},
\Eb{j}, \hc, \hd \brcurl$, where $j$ is an integer, with commutation 
relations given by
\begin{equation}\label{commutation}
\begin{split}
&\blb \H{k}, \H{j} \brb = 2k\d_{k+j,0}\hc, \quad
\blb \H{k}, \Eab{j} \brb = \pm 2 \Eab{k+j}, \quad
\blb \Ea{k}, \Eb{j} \brb = \H{k+j} + k\d_{k+j,0}\hc, \\
&\blb \hd, T^{j} \brb = jT^{j}, \quad \blb \hc, T^{j} \brb = 0,
\quad \text{where} \quad T^{j} \in \lcurl\H{j},\,\Ea{j},\,\Eb{j}\rcurl. 
\end{split}
\end{equation}
In the construction of integrable models we use
the loop algebra only, achieved by setting $\hc=0$.
The homogeneous gradation $Q=\hd$ yields the grading subspaces
\begin{equation}\label{homgrading}
\alie^{(j)} = \lcurl \H{j}, \Ea{j}, \Eb{j} \rcurl.
\end{equation}
Let us fix the semi-simple element as $E^{(1)} \equiv \H{1}$, yielding 
$\ck^{(j)}=\lcurl\H{j}\rcurl$ and $\cm^{(j)}=\lcurl\Ea{j},\,\Eb{j}\rcurl$.
Therefore, the operator containing the fields must have the form 
\begin{equation}
A^{(1)} = q(x,t)\Ea{1}+r(x,t)\Eb{1},
\end{equation}
where $q$ and $r$ are both fields of the models.
The Lax operator $V$ in \eqref{operators_v} is a sum of 
elements in the form $\D{j}=a_j\Ea{j}+b_j\Eb{j}+c_j\H{j}$, whose 
coefficients will be determined in terms 
of $q$ and $r$ by solving the zero curvature 
equation as described in \eqref{set_equations}.

\subsection{Positive flows}
\label{sec:positive_flows}
The positive flows \eqref{pos_flow} are constructed 
from the zero curvature equation
\begin{equation}
\blb \pa_x + \H{1}+q\Ea{1}+r\Eb{1}, \pa_t + \D{n}+\D{n-1}+\dotsb+\D{1}\brb = 0.
\end{equation}
Now let us consider the first models.

\paragraph{$\mathbf{n=1}$.} Yields the trivial equations 
$q_t=q_x$ and $r_t=r_x$.

\paragraph{$\mathbf{n=2}$.} 
Starting from the $\alie^{(3)}$ projection we have $a_2=c_2q$ and $b_2=c_2r$. 
The $\alie^{(2)}$ projection gives
\begin{equation}
a_1 = c_1 q - \tfrac{1}{2}\pa_x a_2, \qquad
b_1 = c_1 r + \tfrac{1}{2}\pa_x b_2, \qquad
\pa_x c_2 = ra_1 - qb_1,
\end{equation}
from which we have
\begin{equation}
\pa_x\ln c_2 = -\tfrac{1}{2}\pa_x\ln\(1+qr\), \qquad c_2 = \( 1+qr \)^{-1/2}.
\end{equation}
The $\ck$ component of the $\alie^{(1)}$ projection implies 
$\pa_x c_1 = 0$. Choosing $c_1=0$%
\footnote{Without loss of generality because it contributes with a linear 
term that can be absorbed by a simple change of coordinates.} 
the $\cm$ component yields the following equations of motion
\begin{equation} \label{wki2}
\pa_t q + \pa_x^{2}\(\dfrac{q}{2\(1+qr\)^{1/2}}\)=0,
\qquad
\pa_t r - \pa_x^{2}\(\dfrac{r}{2\(1+qr\)^{1/2}}\)=0.
\end{equation}
From the coefficients $a_i$, $b_i$ and $c_i$ just found 
we have the explicit Lax pair for this model which reads
\begin{subequations}\label{lax_wki2}
\begin{align}
U &= \H{1} + q\Ea{1}+r\Eb{1}, \\
V &= 
\dfrac{q}{\(1+qr\)^{1/2}}\Ea{2}+
\dfrac{r}{\(1+qr\)^{1/2}}\Eb{2}+
\dfrac{1}{\(1+qr\)^{1/2}}\H{2} \nnn
&\qquad - \pa_x\(\dfrac{q}{2\(1+qr\)^{1/2}}\)\Ea{1}+
\pa_x\(\dfrac{r}{2\(1+qr\)^{1/2}}\)\Eb{1}.
\label{lax_t2}
\end{align}
\end{subequations}
The choice $r=q^*$, $x\to ix$ and $t\to 2it$ reduces \eqref{wki2} to
the \emph{Wadati-Konno-Ichikawa-Shimizu} equation \cite{Shimizu_new_eq}
\begin{equation}\label{shimizu}
i\pa_tq + \pa_x^2\(\frac{q}{\(1 + |q|^2\)^{1/2}}\) = 0.
\end{equation}

\paragraph{$\mathbf{n=3}$.} 
In this case we get the following model
\begin{equation}\label{wki3}
\pa_t q - \pa_x^{2}\(\dfrac{q_x}{4\(1+qr\)^{3/2}}\)=0,
\qquad 
\pa_t r - \pa_x^{2}\(\dfrac{r_x}{4\(1+qr\)^{3/2}}\)=0,
\end{equation}
whose Lax pair is
\begin{subequations}\label{lax_wki3}
\begin{align}
U &= \H{1} + q\Ea{1}+r\Eb{1}, \\
V &= 
\frac{q}{\(1+qr\)^{1/2}}\Ea{3} + 
\frac{r}{\(1+qr\)^{1/2}}\Eb{3} +
\frac{1}{\(1+qr\)^{1/2}}\H{3} \nnn & \qquad 
-\dfrac{q_x}{2\(1+qr\)^{3/2}}\Ea{2}
+\dfrac{r_x}{2\(1+qr\)^{3/2}}\Eb{2}
+\dfrac{q_x r - q r_x}{4\(1+qr\)^{3/2}}\H{2} \nnn & \qquad 
+\pa_x\(\dfrac{q_x}{4\(1+qr\)^{3/2}}\)\Ea{1}
+\pa_x\(\dfrac{r_x}{4\(1+qr\)^{3/2}}\)\Eb{1}.
\label{lax_wki3_v}
\end{align}
\end{subequations}
Choosing $r=q$ and $t\to-4t$ the model \eqref{wki3} becomes
\begin{equation}\label{elastic_beam2}
\pa_t q + \pa_x^2\(\frac{q_x}{\(1+q^2\)^{3/2}}\) = 0,
\end{equation}
which is the EBE \eqref{elastic_beam}.
Note also that, by choosing $r=1$, $q=v-1$ and $t \to -2 t$, 
the system \eqref{wki3} is transformed into the \emph{Dym} equation
\begin{equation}\label{hde}
v_t = \left(v^{-1/2}\right)_{xxx}.
\end{equation}
The Lax pair for the Dym equation, and also for \eqref{elastic_beam2}, are 
obtained from \eqref{lax_wki3} as particular cases. 
The models \eqref{wki2} and \eqref{wki3} were
proposed in \cite{Wadati_wki} and are the first members of the positive 
WKI hierarchy.

\subsection{Negative flows}
\label{sec:negative_flows}
We now extend the WKI hierarchy to incorporate negative flows.
According to our construction, the negative 
flows are generated using the operator \eqref{neg_flow} yielding
\begin{equation}\label{zero_curv_neg_flow}
\blb \pa_x + \H{1} + q\Ea{1} + r\Eb{1}, 
\pa_t + \D{-n}+\D{-n+1}+\dotsb+\D{0}+\D{1}\brb = 0.
\end{equation}

\paragraph{$\mathbf{n=1}$.} 
The projection 
into $\alie^{(-1)}$ implies that
$a_{-1},\; b_{-1}$ and $c_{-1}$ are all constants%
\footnote{
In fact they can
depend on $t$, but this contributes with an overall 
factor that can be absorbed by a change of coordinates, 
so we assume they are constants. This $t$ dependency will be exploited
to construct nonautonomous models in the following section.
}. The $\alie^{(0)}$ projection
yields
\begin{equation}\label{g0_neg1}
\pa_x a_0 = 2c_{-1}q - 2a_{-1}, \qquad
\pa_x b_0 = - 2c_{-1}r + 2b_{-1}, \qquad
\pa_x c_0 = a_{-1}r - b_{-1}q.
\end{equation}

Let us introduce the operator
\begin{equation}
\pa_x^{-1}f(x) \equiv \int_{-\infty}^{x}f(y)dy.
\end{equation}
We assume that the fields and its derivatives of
any order decay sufficiently fast when $|x|\to \infty$. Under this
condition $\pa_x\pa_x^{-1}f(x)=\pa_x^{-1}\pa_xf(x)=f(x)$.

To eliminate an explicit $x$ dependence in \eqref{g0_neg1}
we set $a_{-1}=b_{-1}=0$ thus
\begin{equation}\label{azero}
a_0=2c_{-1}\pa_x^{-1} q, \qquad b_0=-2c_{-1}\pa_x^{-1} r,
\qquad c_0=\text{const.}
\end{equation}
The projection into $\alie^{(2)}$ implies that $a_1=c_1q$ and $b_1=c_1r$. 
The $\alie^{(1)}$ projection yields
the field equations plus one constraint
\begin{equation}
\pa_t q = \pa_x a_1 + 2a_0 - 2c_0q, \qquad
\pa_t r = \pa_x b_1 - 2b_0 + 2c_0r, \qquad
\pa_x c_1 = a_0r - b_0q.
\end{equation}
From this last equation and \eqref{azero} we have
\begin{equation}
c_1 = 2c_{-1}\pa_x^{-1}\(r\pa_x^{-1}q+q\pa_x^{-1}r\). 
\end{equation}
Choosing $c_0=0$\footnote{It contributes with a linear
term that can be absorbed by change of coordinates.} 
and fixing $c_{-1}=1$, we obtain the nonlocal equations of motion
\begin{subequations}\label{wki_nloc}
\begin{align}
\pa_t q &= 
4\pa_x^{-1}q+
2\pa_x\(q\pa_x^{-1}\(r\pa_x^{-1}q+q\pa_x^{-1}r\)\), \\
\pa_t r &= 
4\pa_x^{-1}r+
2\pa_x\(r\pa_x^{-1}\(r\pa_x^{-1}q+q\pa_x^{-1}r\)\).
\end{align}
\end{subequations}
Introducing the new fields defined by
\begin{equation}
q \equiv \pa_x u, \qquad r \equiv \pa_x v,
\end{equation}
we can write \eqref{wki_nloc} in a local form yielding
\begin{subequations}\label{2spe}
\begin{align}
u_{xt} &= 4u + 2\pa_x\(uvu_x\), \label{2spe_1} \\
v_{xt} &= 4v + 2\pa_x\(uvv_x\). \label{2spe_2}
\end{align}
\end{subequations}
This model was also proposed recently in \cite{Matsuno_nspe}%
\footnote{
Choosing $u\equiv\bar{u}+\a\bar{v}$ and
$v\equiv\bar{u}-\a\bar{v}$ where $\a=\sqrt{\mp1}$, after summing and 
subtracting the equations we get 
$\tfrac{1}{4}\bar{u}_{xt} = 
\bar{u}+\tfrac{1}{2}\pa_x\(\(\bar{u}^2\pm\bar{v}^2\)\bar{u}_x\)$ and
$\tfrac{1}{4}\bar{v}_{xt} = 
\bar{v}+\tfrac{1}{2}\pa_x\(\(\bar{u}^2\pm\bar{v}^2\)\bar{v}_x\)$.
Under a small amplitude limit $v \ll 1$ we have
$\tfrac{1}{4}\bar{u}_{xt} = \bar{u}+\tfrac{1}{6}\(\bar{u}^3\)_{xx}$ and
$\tfrac{1}{4}\bar{v}_{xt} = \bar{v}+\tfrac{1}{2}\pa_x\(\bar{u}^2\bar{v}_x\)$,
which were recently considered in \cite{Brunelli_2012} and represents
the small field $\bar{v}$ interacting with the background field $\bar{u}$, 
being a free solution of the SPE.
}.
Note that \eqref{2spe} is symmetric by the interchange of $u$ and $v$ 
and invariant under the following transformations
\begin{equation}
x\to \mu x, \qquad t \to \mu^{-1}t, \qquad u \to \mu^n u, \qquad 
v \to \mu^{2-n}v,
\end{equation}
for a constant $\mu$. The explicit Lax pair of model \eqref{2spe} is given by
\begin{subequations}
\begin{align}
U &= \H{1}+u_x\Ea{1}+v_x\Eb{1}, \\
V &= \H{-1} + 2u\Ea{0} - 2v\Eb{0}+ 2uv\H{1} + 2uvu_x\Ea{1} + 2uvv_x\Eb{1}. 
\label{lax_t-1}
\end{align}
\end{subequations}
If we reduce the model \eqref{2spe} to a single component field by
choosing $v=u$, we have the SPE \eqref{short_pulse}
\begin{equation}\label{spe}
\tfrac{1}{4}u_{xt}  = u + \tfrac{1}{6}\(u^3\)_{xx}.
\end{equation}
Therefore, the model \eqref{2spe} is a two-component generalization of the
SPE and a natural member of the WKI hierarchy, sharing
the same algebraic structure as the previous known models within the positive 
flows. A $n$-component generalization of the SPE, following the structure
of \eqref{2spe}, was also
proposed in \cite{Matsuno_nspe} and our construction  
strongly suggests that it can be obtained by considering the untwisted 
algebra $\hat{A}_{n-1}\sim\hat{s\ell}(n)$, which will also generate
$n$-component extensions of \eqref{shimizu} and \eqref{elastic_beam2}.

\paragraph{$\mathbf{n=2}$.}
Setting $c_{-2}=b_{-2}=0$ to eliminate explicit $x$ dependencies and 
introducing new fields defined through $q\equiv u_{xx}$ and $r\equiv v_{xx}$, 
we get the following model
\begin{subequations}\label{wki_neg2}
\begin{align}
u_{xxt} &= 2a\lcurl 2u_x+\pa_x\(u_xv_xu_{xx}\)\rcurl \nnn
&\qquad - 4b \lcurl  2u + u_xv_xu_{xx} 
+ \pa_x\(u_{xx}\(uv_x-vu_x\)\) \rcurl,
\\
v_{xxt} &= 2a \lcurl 2v_x+\pa_x\(u_xv_xv_{xx}\) \rcurl \nnn
&\qquad +4b \lcurl 2v + u_xv_xv_{xx}
-\pa_x\(v_{xx}\(uv_x-vu_x\)\) \rcurl,
\end{align}
\end{subequations}
where $a\equiv c_{-1}$ and $b\equiv c_{-2}$ are constants. 
The Lax pair for this model is
\begin{subequations}
\begin{align}
U &= \H{1} + u_{xx}\Ea{1} + v_{xx}\Eb{1}, \\
V &= b\H{-2} + 2b u_x\Ea{-1}-2b v_x\Eb{-1} + a\H{-1} \nnn
&\qquad + 2\(a u_x - 2b u\)\Ea{0} - 2\(a v_x+2b v\)\Eb{0}+2b u_xv_x\H{0}
\nnn
& \qquad +2u_{xx}\(a u_xv_x-2b\(uv_x-vu_x\)\)\Ea{1}
+2v_{xx}\(a u_xv_x-2b\(uv_x-vu_x\)\)\Eb{1} \nnn
& \qquad +2\(a u_xv_x-2b\(uv_x-vu_x\)\)\H{1}.
\end{align}
\end{subequations}
Note that the term next to the $a$ coefficient in \eqref{wki_neg2} is  
the system \eqref{2spe}. Setting $a=0$ and
$b=\tfrac{1}{4}$ we have
\begin{subequations}
\begin{align}
\(u_{xt}+u_{xx}\(uv_x-vu_x\)\)_x &= - 2u - u_xv_xu_{xx}, \\
\(v_{xt}+v_{xx}\(uv_x-vu_x\)\)_x &= 2v + u_xv_xv_{xx},
\end{align}
\end{subequations}
which after the substitution $v=u^{*}$ and $t\to i t$ becomes
the single equation
\begin{equation}
\(iu_{xt}+u_{xx}\(u^*u_x-uu^*_x\)\)_x = 2u + u_xu^*_xu_{xx}.
\end{equation}

\subsection{Small amplitude limit}
Let us give a physical motivation by considering a small 
amplitude limit. Consider the reduced second positive flow \eqref{shimizu}.
Assuming that $q$ is small we have 
\begin{equation}
\(1+|q|^2\)^{-1/2}=1-\dfrac{1}{2}|q|^2+\dotsm
\end{equation}
and then \eqref{shimizu} becomes
\begin{equation}\label{shimu}
iq_t + q_{xx} - \tfrac{1}{2}\pa_x^2\(|q|^2q\) + \mathcal{O}\(q^5\)= 0.
\end{equation}
%
Consider a standard multiple scale expansion ($\epsilon \ll 1$)
\begin{equation}\label{multiple_scale}
q = \sum_{n=1}^{\infty}\epsilon^n q_n(x_0,t_0,x_1,t_1,\dotsc),
\qquad x_i = \epsilon^nx, \qquad t_i = \epsilon^nt.
\end{equation}
Taking order of $\epsilon$ in \eqref{shimu} we obtain the 
plain wave solution
\begin{equation}\label{linear_sol}
q_1 = Ae^{i\sigma}+A^*e^{-i\sigma}, \qquad
\sigma = \k x_0 - \k^2t_0, \qquad A=A(x_1,t_1,\dotsc).
\end{equation}
Taking order of $\epsilon^2$, the elimination of the secularity requires
that $A=A(x_1-2\k t_1,x_2,t_2,\dotsc)$ and we
choose $q_2=0$. Finally, the lower nonlinear contribution comes from 
the order of $\epsilon^3$ and after changing to a coordinate system moving 
with the envelope $A$, $\xi_i \equiv x_i-2\k t_i$ and
$\tau_i \equiv t_i$, we obtain the NLS equation
\begin{equation}\label{multi_nls}
i\pa_{\tau_2}A + \pa^2_{\xi_1}A + \dfrac{3\k^2}{2}|A|^2 A = 0.
\end{equation}
Up to this order the small amplitude solution reads
\begin{equation}
q = \epsilon\(Ae^{i\sigma}+A^*e^{-i\sigma}\)+
\epsilon^3\(Be^{3i\sigma}+B^*e^{-3i\sigma}\)+\mathcal{O}\(\epsilon^4\),
\end{equation}
where $A$ is determined from \eqref{multi_nls} and contains a nonlinear
contribution, and $B$ comes from higher order terms. 
The positive flows of 
the WKI hierarchy contain a strong nonlinearity and describes large
amplitude solutions, which in the small amplitude limit recover the
behaviour described by the AKNS models, like the NLS equation.
Due to the large amplitude solutions these models may be of particular
interest in nonlinear optics, plasmas, Bose-Einstein
condensates and water waves.

Performing the same procedure for the EBE \eqref{elastic_beam2}, the 
order of $\epsilon$ gives the linear wave \eqref{linear_sol} with
$\sigma = \k x_0 + \k^3t_0$. The order of $\epsilon^2$ implies
$A=A\(x_1+3\k^2t_1, x_2,t_2,\dotsc\)$ and $q_2=0$. The order of $\epsilon^3$,
after changing to a coordinate system $\xi_i\equiv x_i+3\k^2t_i$ and
$\tau_i\equiv t_i$, gives again the NLS equation
\begin{equation}
i\pa_{\tau_2}A - 3\k\pa_{\xi_1}^2A - 3\k^3|A|^2A = 0.
\end{equation}

Consider the SPE \eqref{spe}.
The $\epsilon$ order gives a linear wave $u_1$ having the same
form as \eqref{linear_sol} with $\sigma=\k x_0-4\k^{-1}t_0$. 
The $\epsilon^2$ order implies
$A=A\(x_1+4\k^{-2}t_1,x_2,t_2,\dotsc\)$ and $u_2=0$. The $\epsilon^3$ order,
in the coordinates $\xi_i\equiv x_i+4\k^{-2}t_i$ and $\tau_i\equiv t_i$, gives
\begin{equation}
i\pa_{\tau_2}A+4\k^{-3}\pa_{\xi_1}^2A+2\k|A|^2A = 0.
\end{equation}
Note that the sine-Gordon equation 
$u_{xt}=\sin u \approx u - \tfrac{1}{6}u^3+\dotsb$ has stronger
nonlinearities than the short pulse equation, but up to order $\epsilon^3$
both of them yields qualitatively the same multiple scale approximation. 
Contrary to the positive flows of the WKI hierarchy,
the short pulse equation does not seem do describe large amplitude solutions.

\subsection{Mixed flows}
\label{sec:mixed}
It is possible to combine a positive flow $n$ with a negative 
flow $m$ by considering the operator \eqref{operators_v}.
Let $n=2$ and $m=1$, leading to the zero curvature equation
\begin{equation}\label{mixed_21}
\blb\pa_x+\H{1}+q\Ea{1}+r\Eb{1},\pa_t+\D{2}+\D{1}+\D{0}+\D{-1}\brb=0.
\end{equation}
We solve each grade projection starting from highest to lowest, exactly
as before, the only modification is that we can let some coefficients, 
that were previously considered constants, now to depend on $t$, thus 
providing the nonautonomous ingredient. 
For individual flows these coefficients were not interesting because they
come as a global factor in the final equation.
Therefore, after solving \eqref{mixed_21}
we obtain
\begin{subequations}\label{wki_21}
\begin{align}
u_{xt} &= -\dfrac{a(t)}{2}\pa_x^2\(\dfrac{u_x}{\(1+u_xv_x\)^{1/2}}\)
+4b(t)\(u+\dfrac{1}{2}\pa_x\(uvu_x\)\), \\
v_{xt} &= +\dfrac{a(t)}{2}\pa_x^2\(\dfrac{v_x}{\(1+u_xv_x\)^{1/2}}\)
+4b(t)\(v+\dfrac{1}{2}\pa_x\(uvv_x\)\), 
\end{align}
\end{subequations}
where $a(t)$ and $b(t)$ are arbitrary functions. Its Lax pair given by
\begin{subequations}
\begin{align}
U &= \H{1}+u_x\Ea{1}+v_x\Eb{1}, \\
V &= a(t)V(+2)+b(t)V(-1),
\end{align}
\end{subequations}
where $V(+2)$ stands for the operator corresponding to the second
positive flow \eqref{lax_t2}, with $q=u_x$ and $r=v_x$, 
and $V(-1)$ is \eqref{lax_t-1}. Model \eqref{wki_21}
is a nonautonomous mixture of \eqref{wki2} and \eqref{2spe}. This model
also admits the reduction $v=u^*$, $t\to it$ and $x\to ix$, yielding
\begin{equation}
u_{xt} = \dfrac{ia(t)}{2}\pa_x^2\(
\dfrac{u_x}{\(1-|u_x|^2\)^{1/2}}\)
-4b(t)\(u-\dfrac{1}{2}\pa_x\(|u|^2u_x\)\).
\end{equation}

In the same way, for $n=3$ and $m=1$, we have
\begin{subequations}\label{wki_31}
\begin{align}
u_{xt} &= \dfrac{a(t)}{4}\pa_x^2\(\dfrac{u_{xx}}{\(1+u_xv_x\)^{3/2}}\)
+4b(t)\(u+\dfrac{1}{2}\pa_x\(uvu_x\)\), \\
v_{xt} &= \dfrac{a(t)}{4}\pa_x^2\(\dfrac{v_{xx}}{\(1+u_xv_x\)^{3/2}}\)
+4b(t)\(v+\dfrac{1}{2}\pa_x\(uvv_x\)\),
\end{align}
\end{subequations}
together with its Lax pair
\begin{subequations}
\begin{align}
U &= \H{1}+u_x\Ea{1}+v_x\Eb{1}, \\
V &= a(t)V(+3)+b(t)V(-1).
\end{align}
\end{subequations}
Choosing $v=u$, \eqref{wki_31} leads to a combination of EBE and SPE
\begin{equation}
u_{xt} = \dfrac{a(t)}{4}\pa_x^2\(\dfrac{u_{xx}}{\(1+(u_x)^2\)^{3/2}}\)
+4b(t)\(u+\dfrac{1}{6}\(u^3\)_{xx}\).
\end{equation}

It is possible to combine any positive flow with a negative one, generating
nonautonomous models inside the hierarchy.
Due to $a(t)$ and $b(t)$ the dispersion relation will have a time 
depend velocity and the solitons will accelerate.
This models may be nice candidates in applications having accelerated 
ultra-short optical pulses \cite{Leblond_mkdv_sg}.

\section{Conservation laws}
\label{sec:conservation}
Consider the linear spectral problem associated to the WKI hierarchy
\begin{equation}\label{spectral_wki}
\(\pa_x + U\) \Psi = 0, \qquad \(\pa_t + V \)\Psi = 0,
\end{equation}
where $U=\H{1}+q\Ea{1}+r\Eb{1}$ and $V$ depends on the particular flow.
Using a matrix representation we have
\begin{equation}
U = \begin{pmatrix}
\l   & \l q \\
\l r & -\l
\end{pmatrix}, \qquad
V = \begin{pmatrix}
V_{11}(\l) & V_{12}(\l) \\
V_{21}(\l) & V_{22}(\l)
\end{pmatrix}, \qquad
\Psi = \begin{pmatrix}
\psi_1 \\
\psi_2
\end{pmatrix},
\end{equation}
where $\l$ is the loop algebra spectral parameter. 
From \eqref{spectral_wki} we write the Riccati form of the spectral 
problem \cite{Wadati_conservation}, 
whose compatibility yields the conservation laws
\begin{align}
\pa_t\(q\Gamma\) - \pa_x\(\l^{-1}\(V_{11} + V_{12}\Gamma\)\) &= 0, 
\label{cons_law1} \\
\pa_t\(r\Gamma^{-1}\) - \pa_x\(\l^{-1}\(V_{22} + V_{21}\Gamma^{-1}\)\) &= 0,
\label{cons_law2}
\end{align}
where we have defined
\begin{equation}\label{gamma}
\Gamma \equiv \dfrac{\psi_2}{\psi_1}.
\end{equation}
Therefore, we can use $q\Gamma$ and $r\Gamma^{-1}$ to construct an 
infinite number of conserved charges assuming a power series in $\l$. 
Deriving $\Gamma$ and $\Gamma^{-1}$ with respect to $x$ and writing
the result in terms of $F \equiv q \Gamma$ and $G\equiv r\Gamma^{-1}$,
we obtain the Riccati equations
\begin{align}
q\pa_x\(\dfrac{F}{q}\) + \l qr - 2\l F - \l F^2 &= 0, \label{riccati} \\
r\pa_x\(\dfrac{G}{r}\) + \l qr + 2\l G - \l G^{2} &= 0, \label{riccati2}
\end{align}
which are the generating equations for the conserved densities. Both
equations lead to the same results, the densities obtained from
\eqref{riccati2} are the same as those from \eqref{riccati} with
the interchange $q\leftrightarrow r$, so we will focus on \eqref{riccati} only.

\subsection{Local charges}
Let us consider the expansion
\begin{equation}\label{local_expansion}
F = \sum_{n=0}^{\infty}f_n \l^{-n}.
\end{equation}
Substituting \eqref{local_expansion} into \eqref{riccati} and taking powers 
of $\l$ we determine each $f_n$ recursively, the first ones are given by
\begin{align}
f_0 &= -1 + \(1+qr\)^{1/2}, \\
f_1 &= -\dfrac{1}{2}\pa_x \ln\(\dfrac{q}{\(1+qr\)^{1/2}}\) +
\dfrac{q_x}{2q\(1+qr\)^{1/2}}, \\
& \ \, \vdots \nn
\end{align}
The charges associated to these densities, which are conserved thanks 
to \eqref{cons_law1}, are
\begin{align}
H_0 &= \int_{-\infty}^{\infty}\(1+qr\)^{1/2}dx, \label{ham_0} \\
H_1 &= \int_{-\infty}^{\infty}\dfrac{q_x}{2q\(1+qr\)^{1/2}}dx, \label{ham_1} \\
& \ \, \vdots \nn
\end{align}
These charges coincide with those in \cite{Wadati_wki} and are the 
Hamiltonians generating the models within the positive flows.

\subsection{Nonlocal charges}
The most general expansion compatible with the Riccati 
equation \eqref{riccati}, able to generate nonlocal densities, is given by
\begin{equation}\label{nonlocal_expansion}
F = \sum_{n=-1}^{\infty}f_{-n}\l^{n}.
\end{equation}
Substituting \eqref{nonlocal_expansion} into \eqref{riccati} and taking the 
lowest order, $\l^{-1}$, we obtain the Bernoulli differential equation
\begin{equation}\label{bernoulli}
\pa_x f_1 -\(\pa_x \ln q\)f_1 = f_1^2,
\end{equation}
whose solution is
\begin{equation}
f_1 = -\pa_x \ln\( \pa_x^{-1}q\).
\end{equation}
Note that it is a total derivative and therefore contributes with a trivial 
charge. The next orders in $\l$ yield the following equations
\begin{align}
\pa_x f_{0} - \(\pa_x\ln q+2f_1\)f_0 &= 2f_1, \label{den0}\\
\pa_xf_{-1} - \(\pa_x\ln q+2f_1\)f_{-1} &= f_0^2 + 2f_0 - qr,\label{den1}\\
\pa_xf_{-2} - \(\pa_x\ln q+2f_1\)f_{-2} &= 2f_0f_{-1} + 2f_{-1},\label{den2}\\
\pa_xf_{-3} - \(\pa_x\ln q+2f_1\)f_{-3} &= f_{-1}^2+2f_0f_{-2}+2f_{-2},
\label{den3} \\
& \ \, \vdots \nn
\end{align}
Integrating \eqref{den0} we obtain
\begin{equation}
f_0 = 2\pa_x\(\dfrac{\pa_x^{-2}q}{\pa_x^{-1}q}\) - 2,
\end{equation}
which again generates a trivial charge. From \eqref{den1} we
have
\begin{equation}
f_{-1} = \dfrac{q}{\(\pa_x^{-1}q\)^2}\[
4\pa_x^{-1}\(q\(\dfrac{\pa_x^{-2}q}{\pa_x^{-1}q}\)^2\)
-4\pa_x^{-3}q - \pa_x^{-1}\(r\(\pa_x^{-1}q\)^2\)
\],
\end{equation}
which is extremely nonlocal and the next densities becomes even more 
cumbersome. At this point we should mention that there is in fact
three sets of nonlocal charges. 
The first set of charges comes from continuing this procedure, 
integrating \eqref{den2} and so on. 
The second set of charges comes from choosing a trivial solution of the 
Bernoulli equation, $f_1=0$, and the solution
$f_0=q$ of \eqref{den0}. The third set of charges comes from the trivial 
choices $f_1=0$ and $f_0=0$. All the subsequent densities are determined 
by the initial choices of $f_1$ and $f_0$.
Therefore, let $f_1=0$ and $f_0=q$, then we have
\begin{align}
f_0 &= q, \\
f_{-1} &= q\pa_x^{-1}\(q-r+2\), \\
f_{-2} &= 2q\pa_x^{-1}\((1+q)\pa_x^{-1}(q-r+2)\), \\
f_{-3} &= q\pa_x^{-1}\(q\(\pa_x^{-1}(q-r+2)\)^2\) \nnn
& \qquad +4q\pa_x^{-1}\((1+q)\pa_x^{-1}\((1+q)\pa_x^{-1}(q-r+2)\)\), \\
& \ \, \vdots \nn
\end{align}
Let $f_1=0$ and $f_0=0$, then
\begin{align}
f_{-1} &= -q\pa_x^{-1}r, \\
f_{-2} &= -2q\pa_x^{-2}r, \\
f_{-3} &= -4q\pa_x^{-3}r+q\pa_x^{-1}\(q\(\pa_x^{-1}r\)^{2}\), \\
& \ \, \vdots \nn
\end{align}
The respective conserved charges are given by
\begin{equation}
H_{-n} = \int_{-\infty}^{\infty}f_{-n}dx, \qquad n=1,2,3,\dotsc
\end{equation}
Let us emphasize that the charges are conserved for the
whole hierarchy of equations, as can be explicitly checked for some of 
the individual models, either from the positive or negative flows.

\section{Gauge and reciprocal transformations}
\label{sec:dressing}
It was proved \cite{Schafer_spe} that \eqref{spe}
does not have real valued, smooth, soliton like solutions in the form 
$u = u(x+ct)$.
The models within the WKI hierarchy possess the loop-soliton kind of 
solutions, and in this section we aim to explain the algebraic origin of this
peculiarity.

\subsection{Dressing method}
First of all, we will show that the dressing method 
\cite{Babelon_dressing}, well known to 
construct soliton solutions of integrable hierarchies, is unable to solve the 
WKI hierarchy.

The dressing method reconstructs the general operator 
$U=E^{(1)}+A^{(0)}[\phi]$ by gauging a vacuum solution $U_0=E^{(1)}$, 
corresponding to the trivial field configuration $\phi=0$. There
must exist \emph{two} dressing operators $\Thpm$ effecting
the following gauge transformations
\begin{equation}\label{dress}
U = \Thpm U_0 \Thpm^{-1} - \(\pa_x\Thpm\)\Thpm^{-1}.
\end{equation}
From \eqref{spectral_wki} we see that there is a gauge freedom for the operator
$U=-\(\pa_x\Psi\)\Psi^{-1}$, i.e. $\Psi \mapsto \Psi h$ where $h$ is a 
constant group element. Consider the spectral problem for the vacuum 
$\(\pa_x + U_0\)\Psi_0 = 0$. Then, the transformations
$\Psi_0 \mapsto \Thp \Psi_0$ and $\Psi_0 \mapsto \Thm \Psi_0$ will reconstruct 
the same operator $U$ if they 
are gauge equivalent, i.e. $\Thp\Psi_0 = \Thm\Psi_0 h $. In other words, 
the relations \eqref{dress} are valid, if and only if, the dressing
operators are related through a Riemann-Hilbert problem
\begin{equation}\label{states}
\Thm^{-1}\Thp = \Psi_0 h \Psi_0^{-1}.
\end{equation}
The dressing operators are further factorized by a Gauss decomposition
\begin{equation}
\Thp = e^{X^{(0)}}e^{X^{(1)}}\dotsm, \qquad
\Thm = e^{Y^{(0)}}e^{Y^{(-1)}}\dotsm,
\end{equation}
where $X^{(j)},Y^{(j)}\in \alie^{(j)}$. We emphasize that \eqref{states} is
essential to construct explicit solutions, and this relation is valid only
upon the existence of \emph{both} transformations $\Thpm$. 

Now consider the WKI operator replaced into \eqref{dress}
\begin{equation}\label{wki_dress}
\H{1}+q\Ea{1}+r\Eb{1} = \Thpm \H{1} \Thpm^{-1} - \(\pa_x\Thpm\)\Thpm^{-1}.
\end{equation}
Taking this transformation with $\Thp$ we see that $X^{(0)}=0$ is a 
solution of the $\alie^{(0)}$ projection. The $\alie^{(1)}$ projection 
yields
\begin{equation}
q\Ea{1}+r\Eb{1} = -\pa_x e^{X^{(1)}} e^{-X^{(1)}},
\end{equation}
which determines $X^{(1)}$ in terms of $q$ and $r$. We can solve
the higher grade projections recursively in terms of the previous terms to 
conclude that, in principle, the problem is solvable. This means 
that $\Thp$ is able to reconstruct $U$. However, this is not enough. The
transformation \eqref{wki_dress} with $\Thm$ still needs to work. Considering
its $\alie^{(1)}$ projection we have
\begin{equation}\label{eq_g1}
\H{1}+q\Ea{1}+r\Eb{1} = e^{Y^{(0)}}\H{1}e^{-Y^{(0)}}.
\end{equation}
Assuming the group parametrization
\begin{equation}\label{b_group}
B \equiv e^{Y^{(0)}} = e^{\chi\Eb{0}}e^{\phi\H{0}}e^{\psi\Ea{0}}
\end{equation}
we obtain
\begin{equation}\label{incompatible}
\H{1}+q\Ea{1}+r\Eb{1} = \(1+2\chi\psi e^{2\phi}\)\H{1}
-2\psi e^{2\phi}\Ea{1} + \(2\chi+2\psi\chi^2e^{2\phi}\)\Eb{1}.
\end{equation}
This relation implies $\psi=0$ or $\chi=0$, consequently $q=0$ or $r=0$ 
showing that $\Thm$ is unable to reconstruct $U$ and therefore, the dressing
method can not be employed to this case.
Nevertheless, we can explore a gauge relation between the 
WKI and AKNS hierarchies \cite{Ishimori_akns_wki, Rogers_reciprocal}.

\subsection{AKNS solutions}
Consider the AKNS spectral problem in space-time coordinates $(y,s)$
\begin{equation}\label{spectral_akns}
\(\pa_y + F\)\Phi = 0, \qquad 
\(\pa_s + G\)\Phi = 0,
\end{equation}
where $F$ and $G$ are the zero curvature Lax pair. We are
omitting the index $n$ labelling each flow, i.e. $G=G(n)$ and $s=s_n$.
Denoting the AKNS fields by $w$ and $z$, the Lax operator $F$ is given
by
\begin{equation}\label{akns_op}
F = \H{1} + w(y,s)\Ea{0} + z(y,s)\Eb{0}.
\end{equation}
From the pure gauge form 
$w\Ea{0}+z\Eb{0} - \pa_y\nu\hc=-\pa_y e^{X^{(0)}}e^{-X^{(0)}}$, where 
$e^{X^{(0)}}=Be^{\nu\hc}$ and $B$ is given by the Gauss decomposition 
\eqref{b_group}, we obtain the following relations \cite{Aratyn_complex_sg}
\begin{align}
\pa_y\phi &= \chi\pa_y\psi e^{2\phi}, &
\pa_s\phi &= \psi\pa_s\chi e^{2\phi}, \label{akns_relations} \\
w &= -\pa_y\psi e^{2\phi}, &
z &= -\pa_y\chi - \chi^2\pa_y\psi e^{2\phi}.
\end{align}

The fields in $B$ are directly related to the famous tau functions, 
obtained by projecting \eqref{states} between the highest weight states
of the algebra (for more details we refer the reader to
\cite{Iraida_toda, Franca_nvc}). The adjoint relations 
are $\(\Ea{j}\)^{\dagger}=\Eb{-j}$, $\(\H{j}\)^{\dagger}=\H{-j}$ and 
$\hc^{\dagger}=\hc$. The fundamental
states are $\lcurl \ket{\mu_0}, \, \ket{\mu_1}\rcurl$ obeying 
$E_{\pm}^{j}\ket{\mu_i}=0$, $\H{j}\ket{\mu_i}=0$ for $j>0$, $i=0,1$
and also $\hc\ket{\mu_i}=\ket{\mu_i}$,
$\Ea{0}\ket{\mu_i}=0$, $\H{0}\ket{\mu_0}=0$ and $\H{0}\ket{\mu_1}=\ket{\mu_1}$.
Let us define the state $\ket{\mu_2}\equiv\Eb{0}\ket{\mu_1}$,
thus the projection of the LHS of \eqref{states} yields
\begin{equation}
e^{\phi} = \dfrac
{\bra{\m_1}\Thm^{-1}\Thp\ket{\m_1}}
{\bra{\m_0}\Thm^{-1}\Thp\ket{\m_0}}, \qquad
\psi = \dfrac
{\bra{\m_1}\Thm^{-1}\Thp\ket{\m_2}}
{\bra{\m_1}\Thm^{-1}\Thp\ket{\m_1}},\qquad
\chi = \dfrac
{\bra{\m_2}\Thm^{-1}\Thp\ket{\m_1}}
{\bra{\m_1}\Thm^{-1}\Thp\ket{\m_1}}.
\end{equation}
From the RHS of \eqref{states} we define the
tau functions, classified in terms of the group element $h$, through
\begin{equation}\label{tau_func_gen}
\tau_{ij} \equiv \bra{\m_i}\Psi_0 h \Psi_0^{-1} \ket{\m_j},
\qquad i,j=0,1,2.
\end{equation}
Therefore,
\begin{equation}\label{tau}
e^{\phi} = \dfrac{\tau_{11}}{\tau_{00}}, \qquad
\psi = \dfrac{\tau_{12}}{\tau_{11}}, \qquad
\chi = \dfrac{\tau_{21}}{\tau_{11}}.
\end{equation}

The group element $h$ generates soliton solutions when it 
is written in terms of vertex operators in the form
\begin{equation}
h = \prod_{i=1}^{N}\exp\(\Gamma_i\).
\end{equation}
The vertex operators should satisfy eigenvalue equations, defining
the dispersion relations for the hierarchy
\begin{equation}
\[\Gamma_i, yU_0 + s_n V_0(n) \] = f_i(y,s_n)\Gamma_i,
\end{equation}
where $U_0=\H{1}$ and $V_0(n)=\H{n}$ are the vacuum Lax pair.
Thus, upon enforcing the nilpotency property of the vertices between the
states, i.e. $\bra{\mu_k}\(\Gamma_i\)^2\ket{\mu_l}=0$, we have the 
explicit form given by
\begin{equation}\label{tau_explicit} 
\begin{split}
\tau_{kl} &= \bra{\mu_k}\prod_{i=1}^N\(1+e^{f_i}\Gamma_i\)\ket{\mu_l} \\
&= \delta_{kl} + \langle \Gamma_1\rangle_{kl} e^{f_1} + \dotsb 
+\langle\Gamma_1\Gamma_2\rangle_{kl}e^{f_1+f_2} + \dotsb 
+\langle\Gamma_1\Gamma_2\Gamma_3\rangle_{kl}e^{f_1+f_2+f_3} + \dotsb
\end{split}
\end{equation}
The relations \eqref{tau} are very important since they relate a general
multi-soliton solution specified by \eqref{tau_func_gen} and 
\eqref{tau_explicit},  and in the following sections
we will write all the results in terms of $\phi$, $\psi$ and $\chi$.

For the AKNS hierarchy we have two vertex operators
\begin{equation}
\Gamma_i \equiv \sum_{j=-\infty}^{\infty}\dfrac{1}{\(\k_i\)^{j}} \Eb{j}, \qquad
\Gamma'_i \equiv \sum_{n=-\infty}^{\infty}\dfrac{1}{\(\k_i\)^{j}} \Ea{j},
\end{equation}
where $\k_i$ is a complex parameter. These vertices satisfy the 
following eigenvalue equations
\begin{subequations}\label{dispersion}
\begin{align}
\[\Gamma_i, y U_0  + s_n V_0(n)  \] &= +2\(\k_i y + \(\k_i\)^n s_n\) \Gamma_i
\equiv \eta_i(y,s_n) \Gamma_i, \\
\[\Gamma'_i, y U_0 + s_n V_0(n) \] &= -2\(\k_i y +\(\k_i\)^n s_n\) \Gamma_i
\equiv \xi_i(y,s_n) \Gamma_i,
\end{align}
\end{subequations}
where $n\in\integer$ correspond to a given flow of the hierarchy 
and we have defined the
dispersion relations $\eta_i$ and $\xi_i$.
For instance, considering $h=e^{\Gamma_1}e^{\Gamma'_2}$ we obtain
the most simple nontrivial solution
\begin{equation}\label{taufunc}
\begin{split}
\tau_{00} = 1 + \dfrac{\k_1\k_2}{(\k_1-\k_2)^{2}}e^{\eta_1+\xi_2}, \qquad&
\tau_{12} = e^{\xi_2}, \\
\tau_{11} = 1 + \dfrac{(\k_2)^2}{(\k_1-\k_2)^{2}}e^{\eta_1+\xi_2}, \qquad&
\tau_{21} = e^{\eta_1}.
\end{split}
\end{equation}
Note that the tau functions have the same form for every model 
withing the hierarchy, only the power $\(\k_i\)^n$ in 
the dispersion relation \eqref{dispersion} changes for different flows.

\subsection{Gauge transformation}
Through a gauge transformation
$\Phi \mapsto \tPhi=g\Phi$ the operator \eqref{akns_op} is transformed into
\begin{equation}\label{gauge}
\tF = g\(\H{1} + w\Ea{0} + z\Eb{0}\)g^{-1} - \pa_y g g^{-1}.
\end{equation}
Let us assume that $g$ is a solution of
\begin{equation}
w\Ea{0} + z\Eb{0} = g^{-1}\pa_y g,
\end{equation}
thus $g=B^{-1}$ from \eqref{b_group} and the transformation 
\eqref{gauge} yields
\begin{equation}\label{almost_wki}
\tF = g\H{1}g^{-1} = \(1+2\chi\psi e^{2\phi}\)\H{1}+
\(2\psi+2\chi\psi^2e^{2\phi}\)\Ea{1} - 2\chi e^{2\phi}\Eb{1}.
\end{equation}
This operator is not in the form of the WKI operator
\begin{equation}\label{wki_op}
U = \H{1} + q(x,t)\Ea{1} + r(x,t)\Eb{1}
\end{equation}
due to the coefficient of $\H{1}$. If we impose an equality between 
\eqref{wki_op} and \eqref{almost_wki} we fall into the same kind of
inconsistency encountered in \eqref{incompatible}, implying $\psi=0$ or 
$\chi=0$ and then $q=0$ or $r=0$. The correct way to ``normalize'' the
operator \eqref{almost_wki}, absorbing the coefficient of $\H{1}$, is
through a reciprocal transformation.

\subsection{Reciprocal B\" acklund transformation} 
Let us use the results from \cite{Rogers_reciprocal} which asserts that, the 
continuity equation
\begin{equation}\label{conservation1}
\pa_{s} \a(y,s) + \pa_y \b(y,s) = 0,
\end{equation}
is transformed to the reciprocally associated equation
\begin{equation}\label{conservation2}
\pa_{t} \g(x,t) + \pa_x \d(x,t) = 0,
\end{equation}
by the transformation $(y,s)\mapsto(x,t)$ such that
\begin{equation}\label{reciprocal1}
dx = \a dy - \b ds, \qquad dt = ds,
\end{equation}
together with its inverse
\begin{equation}\label{reciprocal2}
dy = \g dx - \d dt, \qquad ds = dt,
\end{equation}
provided that
\begin{equation}\label{reciprocal_relations}
\g = \dfrac{1}{\a}, \qquad \d = -\dfrac{\b}{\a}.
\end{equation}
This theorem is straightforward to prove, noting 
that \eqref{conservation1} and \eqref{conservation2} are consequences
of the compatibility of mixed second derivatives of
\eqref{reciprocal1} and \eqref{reciprocal2}, respectively, while 
\eqref{reciprocal_relations} ensures the reciprocity between both
coordinate systems.

\begin{figure}
\begin{center}
\includegraphics[width=.46\textwidth]{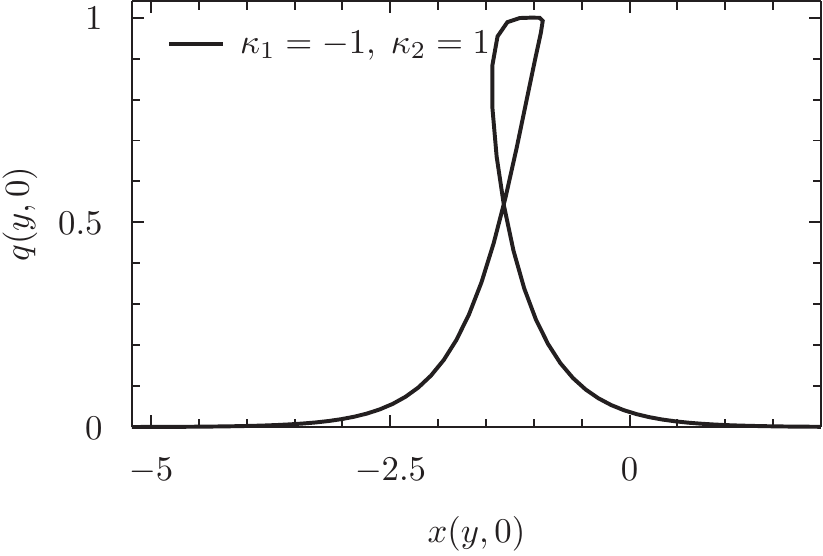}\hspace{2em}
\includegraphics[width=.46\textwidth]{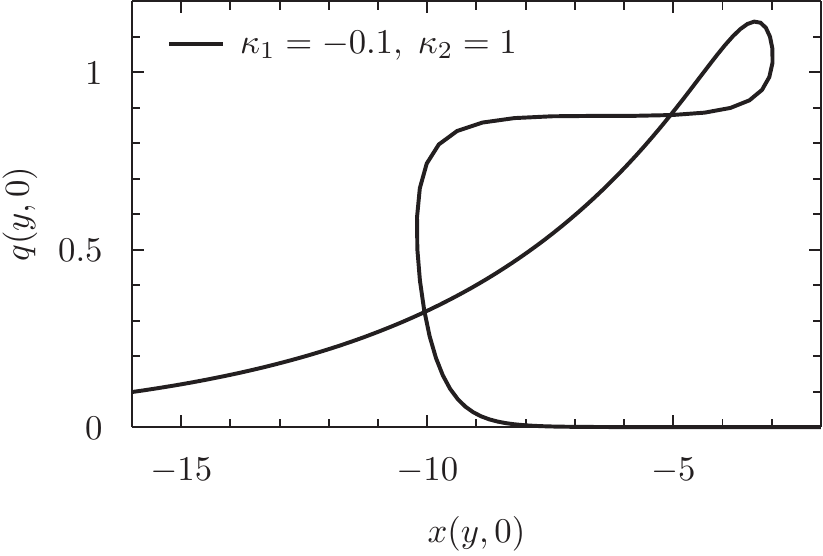}
\caption{Parametric plots of $q(y,t)$ versus $x(y,t)$ at instant $t=0$.
Similar behaviour is obtained for $r(y,t)$. The solutions keep their form
unchanged during $t$ evolution.
\label{fig1}}
\end{center}
\end{figure}

Therefore, using the transformation \eqref{reciprocal1} 
with the operator \eqref{almost_wki} at hands, we get the following
spectral problem in the coordinates $(x,t)$
\begin{equation}\label{spec2}
\( \pa_x + U \)\tPhi = 0, \qquad
\( \pa_{t} + V \) \tPhi = 0,
\end{equation}
where
\begin{equation}\label{ops2}
U = \dfrac{1}{\a}\tF, \qquad
V = {\tG} + \dfrac{\b}{\a}\tF.
\end{equation}
From \eqref{almost_wki} it is evident that defining $\a$ through
\begin{equation}\label{alfa}
\a(y,s) \equiv 1 + 2\chi\psi e^{2\phi}
\end{equation}
the spectral problem \eqref{spec2} is mapped into the WKI problem with
the relations
\begin{equation}\label{wki_sol}
q(y,s) = \dfrac{2\psi+2\chi\psi^2 e^{2\phi}}{\a}, \qquad
r(y,s) = -\dfrac{2\chi e^{2\phi}}{\a}.
\end{equation}
We still need to express the space-time dependence in the 
variables $(x,t)$, so from
\eqref{reciprocal1} we have the implicit relation
\begin{equation}\label{implicit}
\begin{split}
x(y,t) &= \int \a(y,t)dy + \mbox{const.} \\
 &= y + 2\int\dfrac{\tau_{12}\tau_{21}}{\(\tau_{00}\)^2}dy + \mbox{const.}
\end{split}
\end{equation}
where we have used \eqref{alfa} and \eqref{tau} to express it 
in terms of the tau functions.
Writing also \eqref{wki_sol} in terms of the tau functions,
\begin{equation}\label{wki_tau}
q(y,s) = 2\dfrac{\tau_{12}}{\tau_{11}}
\(\dfrac{\(\tau_{00}\)^2 + \tau_{12}\tau_{21}}{
\(\tau_{00}\)^2+2\tau_{12}\tau_{21}}\), \qquad
r(y,s) = -2\dfrac{\tau_{11}\tau_{21}}{\(\tau_{00}\)^2+2\tau_{12}\tau_{21}}.
\end{equation}
Equations \eqref{wki_tau} and \eqref{implicit} are the general solution of
all the models within the WKI hierarchy in terms of the AKNS tau 
functions, which are known.
For instance, replacing \eqref{taufunc} and dispersion 
\eqref{dispersion} with $n=3$, corresponding to the solutions of model 
\eqref{wki3}, the integral in \eqref{implicit} is elementary 
(we also set the integration constant to zero) and an explicit solution is 
obtained, whose graph is sketched in Fig. \ref{fig1}. The mathematical origin
behind the loop-soliton solution is an implicit relation between the 
coordinates, arising from a reciprocal transformation. Multi-loop-solitons
can be constructed similarly using the general form \eqref{tau_explicit}.

\subsection{From SPE to sine-Gordon}
This section is a particular case of a more general
result presented in the next section, but because the hodograph 
transformation $(y,s)\to(x,t)$ previously introduced 
\cite{Sakovich_spe_integrable,Sakovich_solitary_spe,Matsuno_spe_periodic} 
comes out so naturally from our definition \eqref{alfa}, we will consider it 
first and it will also serve as a motivation for the forthcoming section. 
The model \eqref{spe} is obtained from \eqref{2spe} when $v=u$, 
or equivalently $r=q$, thus imposing this restriction upon \eqref{wki_sol} 
we get
\begin{equation}
\chi = -\dfrac{\psi e^{-2\phi}}{1+\psi^2},
\end{equation}
which substituted in the first relation of \eqref{akns_relations} yields
\begin{equation}\label{psi_chi}
e^{-2\phi} = 1+\psi^2, \qquad \chi = -\psi.
\end{equation}
Therefore, \eqref{alfa}, note also \eqref{reciprocal_relations},  
can be written in terms of one of the fields only
\begin{equation}\label{alfa2}
\a = \g^{-1} = \dfrac{1-\psi^2}{1+\psi^2} = 
\dfrac{1-\chi^2}{1+\chi^2} = 2e^{2\phi}-1.
\end{equation}
Reminding from the trigonometric identity 
$\cos\theta=2\cos^2\tfrac{\theta}{2}-1$, the relation \eqref{alfa2} 
strongly suggests the parametrization 
\begin{equation}\label{trigo}
\a=\g^{-1}=\cos\theta, \qquad e^{\phi}=\cos\dfrac{\theta}{2},
\qquad \psi = -\chi = \tan\dfrac{\theta}{2}.
\end{equation}
The transformation \eqref{reciprocal1} implies $u_x=\a^{-1}u_y$, but
$q=u_x$ so from \eqref{wki_sol} we therefore have
\begin{equation}\label{uy}
u_y = -2\chi e^{2\phi} = 2\tan\tfrac{\theta}{2}\cos^2\tfrac{\theta}{2}=
\sin\theta.
\end{equation}
Squaring $u_x=\a^{-1} u_y$ we have
\begin{equation}\label{mat_rel}
u_x^2=(\cos^2\theta)^{-1}\(1-\cos^2\theta\)=\dfrac{1}{\a^2} - 1, 
\qquad \gamma^2 = 1 + u_x^2,
\end{equation}
which is the relation from where \cite{Matsuno_spe_periodic} started. 
Deriving both
sides of \eqref{mat_rel} and plugging into the equations of motion 
\eqref{2spe} (with $v=u$) we have the conservation law 
\eqref{conservation2} and its reciprocal \eqref{conservation1}, namely
\begin{equation}\label{cons2}
\pa_{t}\g = \pa_x\(2u^2\g\), \qquad \pa_{s}\a = \pa_y\( -2u^2 \).
\end{equation}
Using \eqref{trigo} and \eqref{uy} in the second equation
of \eqref{cons2}, we finally get the sine-Gordon model
\begin{equation}\label{sge}
u=\tfrac{1}{4} \theta_s, \qquad  
\theta_{ys} = 4\sin\theta.
\end{equation}

\subsection{From 2-SPE to Lund-Regge}
Let us remind from the identities $q=u_x$, $r=v_x$, $\a=\tfrac{1}{\g}$ 
and $\pa_x=\a^{-1}\pa_y$. Taking the product of both terms in 
\eqref{wki_sol} yields
\begin{subequations}\label{remarkable}
\begin{align}
\g^2 &= 1 + u_xv_x, \label{gamma_square}\\
\a^2 &= 1-u_yv_y. \label{alfa_square}
\end{align}
\end{subequations}
Deriving \eqref{gamma_square} and replacing into \eqref{2spe_1} one gets
\begin{equation}\label{first}
2\dfrac{\g\g_{t}}{v_x} = \dfrac{u_xv_{xt}}{v_x}+4u+
4\dfrac{uv\g\g_x}{v_x}-2\dfrac{uv u_x v_{xx}}{v_x}+
2u_x^2 v +2u\(\g^2-1\),
\end{equation}
while from \eqref{2spe_2} one has
\begin{equation}\label{second}
\dfrac{u_xv_{xt}}{v_x} = 4\dfrac{u_xv}{v_x}+
2\dfrac{uvu_xv_{xx}}{v_x}+2u_x^2v+2u\(\g^2-1\).
\end{equation}
Simplifying these two last equations we obtain the respective 
conservation law \eqref{conservation1} and its reciprocal 
\eqref{conservation2}, which read
\begin{subequations}
\begin{align}
\pa_{t}\g &= \pa_x\(2uv\g\), \label{conservation3} \\
\pa_{s}\a &= \pa_y\(-2uv\). \label{conservation4}
\end{align}
\end{subequations}
Deriving \eqref{alfa_square} with respect to $s$ and using 
\eqref{conservation4} we therefore have
\begin{equation}
v_y\(u_{ys}-4\a u\) + u_y\(v_{ys} - 4\a v\) = 0.
\end{equation}
A consistency condition, e.g. under the choice $v=u$, requires that
\begin{subequations}\label{double}
\begin{align}
u_{ys} &= 4\a u, \label{double_u} \\
v_{ys} &= 4\a v. \label{double_v}
\end{align}
\end{subequations}
From \eqref{double} we have $u$ and $v$ in terms 
of $\chi, \phi, \psi$, once $u_y=\a q$ and $v_y = \a r$ are
given from \eqref{wki_sol}. 
However, for our purposes, it is more convenient to introduce the auxiliary 
fields
\begin{equation}\label{tilde}
\bpsi \equiv \psi e^{\phi}, \qquad \bchi \equiv \chi e^{\phi} .
\end{equation}
Then, the relations \eqref{akns_relations} read 
\begin{equation}\label{new_relations}
\pa_y\phi = \dfrac{\bchi\pa_y\bpsi}{1+\bchi\bpsi}, \qquad
\pa_{s}\phi = \dfrac{\bpsi\pa_{s}\bchi}{1+\bchi\bpsi},
\end{equation}
and \eqref{wki_sol} yields
\begin{subequations}\label{der_tilde}
\begin{align}
u_y &= 2\bpsi\(1+\bchi\bpsi\)e^{-\phi}, \label{uy_tilde} \\
v_y &= -2\bchi e^{\phi}. \label{vy_tilde}
\end{align}
\end{subequations}
Replacing \eqref{der_tilde} into \eqref{double} we obtain
\begin{subequations}\label{final_tilde}
\begin{align}
u &= \dfrac{\pa_{s}\bpsi e^{-\phi}}{2}, \label{u_tilde} \\
v &= -\dfrac{\pa_{s}\bchi e^{\phi}}{2\(1+\bchi\bpsi\)}, \label{v_tilde}
\end{align}
\end{subequations}
and if we employ \eqref{tilde} with \eqref{tau}, the solution in 
terms of tau functions reads
\begin{subequations}\label{2spe_tau_sol}
\begin{align}
u(y,s) &= 
\dfrac{\tau_{00}\pa_{s}\tau_{12}-\tau_{12}\pa_{s}\tau_{00}}{
2\tau_{00}\tau_{11}},\\
v(y,s) &= -\dfrac{\tau_{11}}{2\tau_{00}}
\dfrac{\tau_{00}\pa_{s}\tau_{21}-\tau_{21}\pa_{s}\tau_{00}}{\tau_{00}^2
+\tau_{12}\tau_{21}}.
\end{align}
\end{subequations}

Deriving \eqref{final_tilde} with respect to $y$ and equating to
\eqref{der_tilde}, after using \eqref{new_relations},
we obtain the Lund-Regge model \cite{Lund_Regge,Lund_classical}
\begin{subequations}\label{lund_regge}
\begin{align}
\bpsi_{ys} - \dfrac{\bchi \bpsi_y \bpsi_s}{1+\bchi\bpsi} - 
4\bpsi\(1+\bchi\bpsi\) &= 0, \\
\bchi_{ys} - \dfrac{\bpsi \bchi_y \bchi_s}{1+\bchi\bpsi}-
4\bchi\(1+\bchi\bpsi\) &= 0.
\end{align}
\end{subequations}
Therefore, we have explicitly established an equivalence between 
the 2-SPE \eqref{2spe} and the Lund-Regge model \eqref{lund_regge}.

As a particular case, choosing $\bchi=-\bpsi$ we have
\begin{equation}\label{sg2}
\bpsi_{ys} + \dfrac{\bpsi \bpsi_y \bpsi_s}{
1-\bpsi^2}-4\bpsi\(1-\bpsi^2\) = 0,
\end{equation}
which is the sine-Gordon equation \eqref{sge} through the parametrization
$\bpsi = \sin\tfrac{\theta}{2}$.

\section{Conclusions}
\label{sec:conclusions}
We have proposed a higher grading construction for integrable
hierarchies. When the algebra $\alie=\hat{A}_1$ with homogeneous gradation
is chosen, this construction yields the WKI hierarchy.
The previous known models \cite{Wadati_wki} constitute 
the positive flows. We have extended the WKI hierarchy to negative flows, where
the first negative flow yields the two-component field generalization
of the short pulse equation \cite{Schafer_spe}. Therefore, it is clear that  
in all these models underlie the same algebraic structure.
Some novel integrable nonautonomous models were also proposed, 
mixing a positive with a negative flow. These integrable mixed models may have 
applications in nonlinear optics, specially concerning accelerated 
ultra-short optical pulses \cite{Leblond_mkdv_sg}.

We have derived \emph{local} and \emph{nonlocal} charges from the Riccati
form of the spectral problem. The nonlocal charges 
are usually more difficult to be obtained, and we have shown how
they can be constructed through positive power series expansion 
in the spectral parameter. This method can be used to other 
integrable hierarchies as well.

We have demonstrated that the dressing method is unable to solve the WKI 
hierarchy, 
which in fact, does not have the usual solitonic type of solution, 
travelling with constant velocity. 
Combining gauge and reciprocal B\" acklund transformations we 
have established the precise connection between the whole WKI and AKNS 
hierarchies, mapping each flow of both hierarchies. 
This gives a formal explanation of various relations scattered through the 
literature, for instance, the connection between the elastic beam  
and mKdV equations \cite{Ishimori_loop_soliton}, the relation between the 
Dym and KdV equations \cite{Hereman_dym_kdv} and also the recent 
hodograph transformation \cite{Sakovich_spe_integrable} relating the short 
pulse equation to the sine-Gordon model. We considered explicitly 
this mapping between the first negative flows of the WKI and AKNS 
hierarchies, and demonstrated the correspondence between the novel
two-component short pulse equation \eqref{2spe} and the Lund-Regge model 
\eqref{lund_regge}, generalizing the previous hodograph 
transformation \eqref{hodograph}. In the same way that
the notorious Miura transformation relates the KdV and mKdV hierarchies,
which in fact is a gauge transformation, the combination of gauge followed
by a reciprocal transformation yields more involved relations.
We believe that our results show this explicitly, and the same approach
can be applied to other integrable hierarchies following our construction. 
This relations will generate exotic solitonic solutions, like the 
loop-solitons, arising from an implicit space-time dependence.

We have constructed solutions for the whole WKI hierarchy, writing
them in terms of the known AKNS tau functions. 
We have motivated that the models 
within the positive flows of the WKI hierarchy describe larger amplitude 
solutions than the corresponding usual solitonic solutions, which is expected 
from the higher nonlinearity contained in the equations of motion. 
We believe that these models may have particular interesting applications 
that were not further explored.

Finally, we stress that our higher grading construction is general and other 
affine Lie algebras can be considered, which will give rise 
to novel integrable models. For instance, a $n$-component generalization
can be considered with the algebra $\hat{A}_{n-1} \sim \hat{s\ell}_n$.
The technique of using gauge plus reciprocal 
transformations will be well suited to deal with these cases.

\subsubsection*{Acknowledgments}
We thank CAPES, CNPQ and Fapesp for financial support.

\bibliographystyle{bibstyle}
\bibliography{biblio}

\end{document}